\documentclass[a4paper]{article}

\usepackage{spconf}
\usepackage{subfig}
\usepackage{amsmath}
\usepackage{booktabs}
\usepackage{graphicx}
\usepackage{url}
\usepackage{mathtools}
\usepackage{enumitem}
\usepackage{fancyhdr}

\fancypagestyle{firstpage}{%
\fancyhf{}
\fancyfoot[C]{\copyright2019 IEEE. Published in the IEEE Automatic Speech Recognition and Understanding Workshop (ASRU),\\14-18 December 2019 in Sentosa, Singapore.}

}

\title{Towards Real-time Mispronunciation Detection in Kids' Speech}
\name{Peter Plantinga, Eric Fosler-Lussier}
\address{The Ohio State University}
\begin{document}

\maketitle
\thispagestyle{firstpage}
\begin{abstract}
  Modern mispronunciation detection and diagnosis systems have seen significant gains in accuracy due to the introduction of deep learning. However, these systems have not been evaluated for the ability to be run in real-time, an important factor in applications that provide rapid feedback. In particular, the state-of-the-art uses bi-directional recurrent networks, where a uni-directional network may be more appropriate. Teacher-student learning is a natural approach to use to improve a uni-directional model, but when using a CTC objective, this is limited by poor alignment of outputs to evidence. We address this limitation by trying two loss terms for improving the alignments of our models. One loss is an ``alignment loss'' term that encourages outputs only when features do not resemble silence. The other loss term uses a uni-directional model as teacher model to align the bi-directional model. Our proposed model uses these aligned bi-directional models as teacher models. Experiments on the CSLU kids' corpus show that these changes decrease the latency of the outputs, and improve the detection rates, with a trade-off between these goals.
\end{abstract}

\section{Introduction}

Reading and pronunciation are essential skills for children to learn. In order to reduce the cost and increase availability, computer-assisted pronunciation training (CAPT) has attracted more research interest recently. While some approaches do not use automatic speech recognition (ASR) \cite{lee2012comparison}, using ASR has been shown to be effective for training children, and merits closer investigation \cite{neri2008effectiveness}.

Deep learning has improved the quality of these systems, but since their introduction, they have not been evaluated for the ability to be run in real-time. This is an important factor for being able to use these systems in production, since real-time feedback can greatly enhance the experience of the user.

In particular, the state-of-the-art for mispronunciation detection and diagnosis (MDD) is a bi-directional recurrent neural network with gated recurrent units (BiGRU) \cite{leung2019cnn} that is trained using connectionist temporal classification (CTC) \cite{graves2006connectionist}. This model has the limitation that it requires both backward and forward contexts in order to make its decision, so it cannot operate without the entire utterance being recorded.

We seek to address this by using uni-directional GRUs (UniGRUs). However, these models have two primary limitations. First, they tend to produce delayed outputs, preferring to wait until all evidence has been seen before making a prediction. Second, the performance of these models is significantly worse than the bi-directional models.

We experiment with using teacher-student learning to improve the performance and latency of the uni-directional models. However, models trained with CTC are not well-suited for teacher-student learning, because the outputs are not aligned with the evidence. To deal with this, we seek to align the outputs with the evidence in the BiGRU, so that it can be used effectively as a teacher model. Our formulation of teacher-student learning for CTC-based models can be summarized with two steps:

\begin{enumerate}
    \item We align the outputs of a BiGRU by training the model with additional loss terms. The two terms that we experiment with are ``alignment loss'' which encourages the blank symbol during silence and non-blank symbols otherwise, and teacher-student learning with a UniGRU as teacher.
    \item We train a new UniGRU model using the aligned BiGRU as the teacher model. These models have improved latency and error rate, with some trade-off between these two metrics of performance.
\end{enumerate}

%I THINK YOU NEED ONE SENTENCE HERE ON WHAT IS NEW RELATIVE TO PREVIOUS STUDENT-TEACHER LEARNING TECHNIQUES.
To our knowledge, our work is one of the first to successfully explore aligning the outputs of a CTC model to the evidence, providing potential benefits for knowledge transfer, ensemble learning, and interpreting outputs. This work is complimentary to other bi-to-uni knowledge transfer methods, allows a broader range of possibilities for accuracy-vs-latency trade-off, and bridges 40\% of the gap in mispronunciation detection performance between bi-directional and uni-directional models.

\section{Prior Work}

Some researchers have made efforts to do CAPT without ASR, given the limitations of the approach. One prominent example is the work done by Ann Lee in \cite{lee2012comparison,lee2013mispronunciation,lee2016personalized}, comparing unsupervised features using dynamic time warping (DTW) and deep belief networks (DBN). In later work, the authors propose a model for unsupervised error pattern discovery.

By contrast, most efforts at CAPT are ASR-based. One of the first examples is the extended recognition network (ERN) \cite{harrison2009implementation} which takes a traditional decoding approach and appends commonly mis-pronounced phones to the graph, allowing decoding to take one of several paths based on the observations. If the ERN produces a sequence of phones that includes an incorrect phone, the word is marked as being incorrectly pronounced.

To overcome the limitations of this approach, researchers have done free-phone recognition using multi-distribution networks \cite{li2016mispronunciation}. This approach, termed APGR, uses acoustic, graphemic, and phonemic representations to decide when a mispronunciation has occurred. The acoustic features are incorporated via a monophone acoustic model, trained with forced-alignments.

Alternative approaches get around the need for forced alignments by using end-to-end models to do MDD, notably the state-of-the-art work on training CNN-RNN-CTC models \cite{leung2019cnn}. Convolutional layers are used to capture higher-level acoustic features, while context is modeled with a BiGRU. Unfortunately the use of forward context by the RNN makes this model unsuitable for real-time applications.

Since uni-directional models, which are more amenable for real-time applications, typically have weaker performance than bi-directional models, it is beneficial to explore ways to close the gap in performance. This has often done using knowledge transfer. Research on knowledge transfer was introduced with the concept of teacher-student learning \cite{ba2014deep}, for the task of model compression. In this formulation, the logits of the model (before softmax) are used as a "soft label" for the student model, giving more information than a single correct answer (e.g. about relative similarity of non-correct answers). A simple mean-square-error (MSE) between the logits of teacher model and student model is used as a loss function.

The knowledge transfer literature was extended with the introduction of knowledge distillation \cite{hinton2015distilling}. This formulation is similar to teacher-student learning, but uses the posterior (after softmax) and cross-entropy loss with a temperature parameter.

There has been some work on transferring knowledge from bi-directional models to uni-directional models, when training with CTC. One approach is to perform decoding and use an n-best list of decoded posteriors as targets \cite{takashima2018investigation}. This approach was introduced by \cite{kim2016sequence} for the task of neural machine translation. The authors followed their work with lattice-based sequence-level knowledge distillation \cite{takashima2019investigation}.

\begin{figure}[t!]
    \centering
    \includegraphics[width=\linewidth]{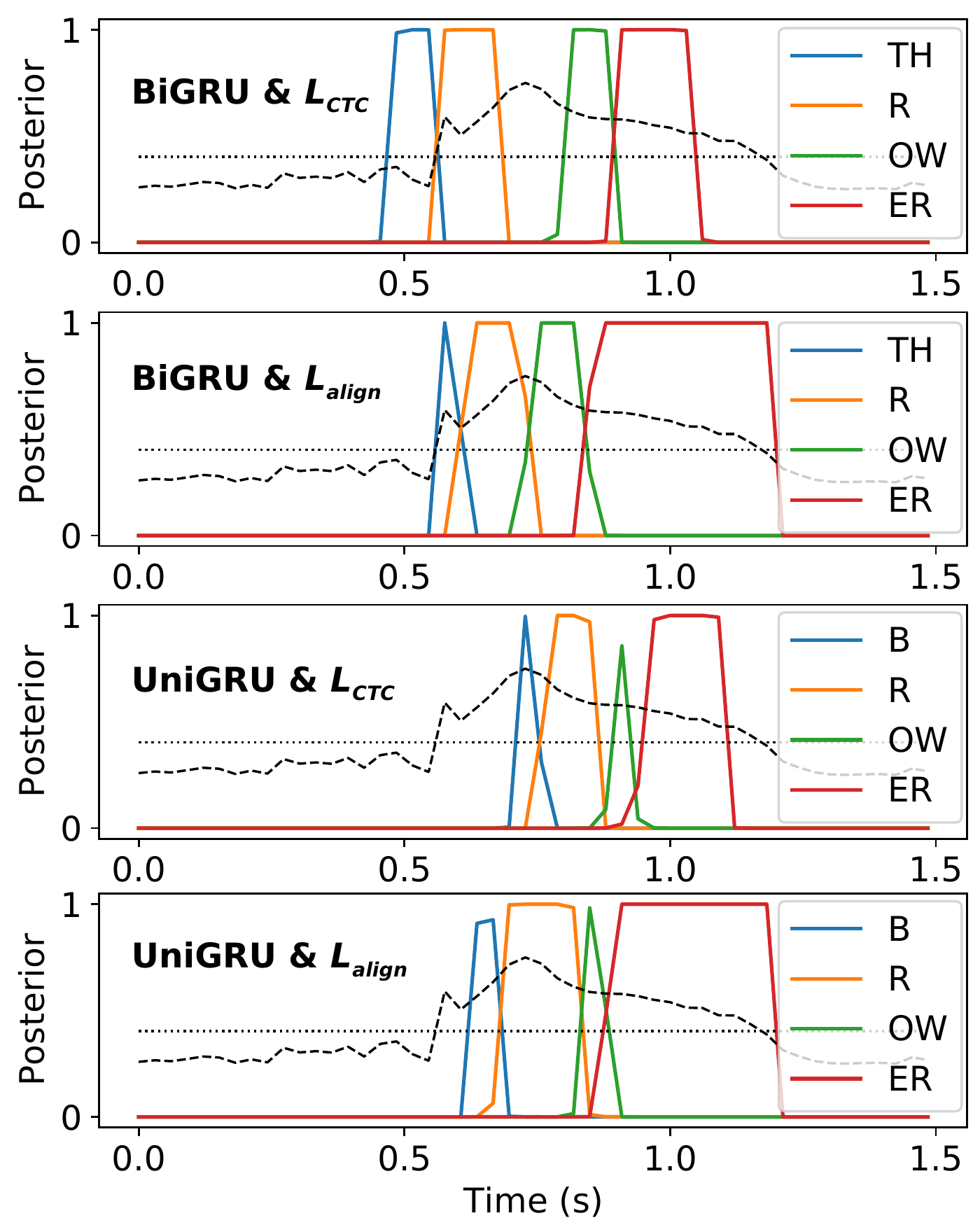}
    \caption{Comparison of before and after the addition of alignment loss for both a BiGRU and a UniGRU. The utterance id is ks1187w0 and the speaker's prompt is "thrower". Listening to the utterance suggests the pronunciation sounds more like "B R AW ER" than "TH R OW ER", though the utterance is labeled as having correct pronunciation. The dashed line represents the average energy of each frame, and the dotted line represents the threshold for alignment loss between frames that are considered "silence" vs. "non-silence".}
    \label{fig:align}
\end{figure}

These authors report that even with extensive optimization, sequence-level knowledge distillation trains 3-5 times slower than frame-level. An alternative suggested by Kurata and Audhkhasi \cite{kurata2018improved} seeks to alleviate the issue of model misalignment by comparing each frame of the student model's output to a window of several previous frames of the teacher model. This approach has the weakness that the frames with the most helpful information might not be within the window of examined frames, since the BiGRU sometimes produces output well before the evidence.

Work on aligning the outputs of end-to-end models includes end-to-end maximum mutual information (EEMMI) loss, introduced in \cite{fritz2017simplified}. This loss requires a language model for both training and decoding, which may be unsuitable for mispronunciation detection, since it may correct errors in pronunciation automatically.

Reducing the latency of uni-directional models was explored in \cite{sak2015acoustic}, where the authors constrain the forward-backward alignment during training using labeled phoneme boundaries. The authors were unable to gain improvement from knowledge transfer.

In the next section, we outline a new teacher model for aligning outputs that is compatible with Kurata and Audhkhasi's work and improves pronunciation error detection.

\section{Aligning the outputs}

We seek to align the outputs of the BiGRU with the evidence, so that this model can serve as an effective teacher model. To do this, we explore two different additions to the loss function, in order to encourage aligned outputs: an "alignment loss" and teacher-student learning, with the UniGRU as the teacher.

\subsection{Alignment loss}

\begin{figure}[t]
    \centering
    \includegraphics[width=\linewidth]{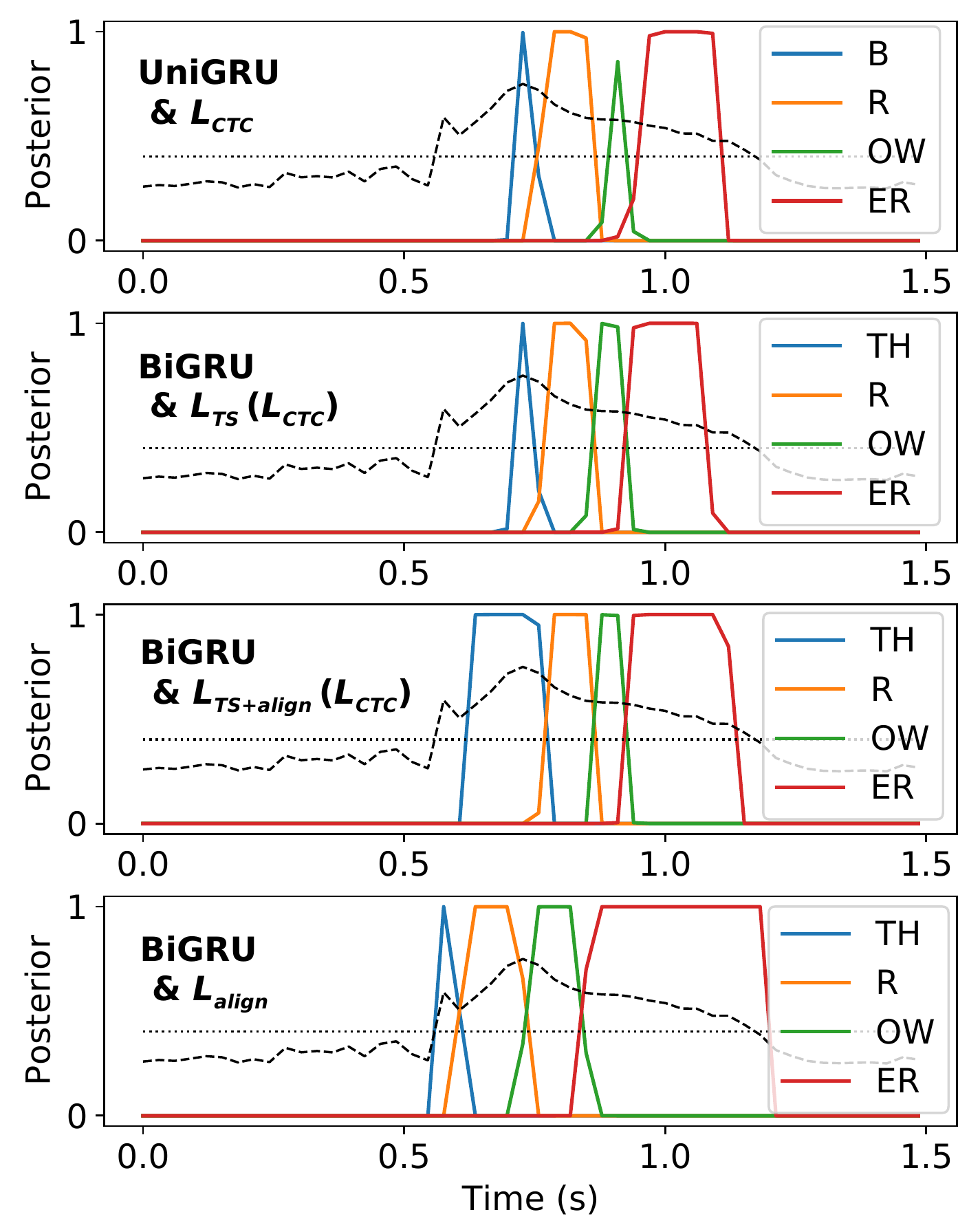}
    \caption{Comparison of BiGRU alignments after adding teacher-student learning. A UniGRU was used as the teacher model.  }
    \label{fig:teacher_align}
\end{figure}

When trained with CTC, both BiGRUs and UniGRUs tend to have outputs that are not aligned with the evidence \cite{kurata2018improved}. In the case of UniGRUs, the outputs tend to be delayed until the model has gathered enough evidence to output a symbol. In the BiGRU case, outputs can occur either before or after the evidence.

One way to encourage better alignments is to penalize outputs that occur during silence, and similarly to penalize the blank symbol during non-silence portions of the utterance. Even a simple definition of silence will encourage better alignments, so we choose to define silence as frames where the average energy of the frame is less than the average energy of the utterance.

Then the loss term is simply the log of the posterior. If the frame is marked silence, we use the posterior not including the blank symbol, to encourage outputting blank symbols, and vice versa for the frames marked non-silence. Our scoring function $f(\cdot)$ is defined by the predicted phoneme $\hat{p}_t$ and energy $E_t$ at frame $t$ (out of $T$), as well as the parameters of the network, $\theta$.

\begin{equation}
    f(t) = \begin{dcases}
        \Pr(\hat{p}_t \neq \epsilon; \theta) & \text{if } E_t < \frac{1}{T} \sum_{i=1}^T E_i \\
        \Pr(\hat{p}_t = \epsilon; \theta) & \text{otherwise} \\[10pt]
    \end{dcases}
\end{equation}

Then our alignment loss is simply the average of the log of the scoring function applied to all frames, in addition to CTC:

\begin{equation}
    L_{align} = L_{CTC} + \frac{1}{T} \sum_{i=1}^T \log(f(i))
\end{equation}

This loss seems to significantly improve alignments, as seen in Figure \ref{fig:align}. In particular, the BiGRU's outputs seem to match the evidence very closely, with each posterior peak's onset aligned with a change in direction of the average energy. The UniGRU's output is still slightly delayed from the evidence, but much more closely aligned than without the additional loss term. Interestingly, the UniGRU seems to find a more accurate phoneme transcript than the BiGRU, perhaps due to lack of forward context with which to guess the "correct" transcript.

In addition to the example, we collect statistics about the outputs and show them in Table \ref{tab:stats}. The alignment loss significantly increases average number of frames that are not blank, going from 28.8 to 37.7 frames in the case of the BiGRU, and from 24.1 to 31.0 frames in the UniGRU.

In terms of the average number of peaks, and their average onset, not much change was observed in the BiGRU, while the UniGRU has an increase in number of peaks and earlier outputs.

\begin{table}
\centering
\begin{tabular}{llccc}
    \toprule
    Model & Loss & \# frames & \# peaks & delay  \\
    \midrule
    BiGRU & $L_\textit{CTC}$ & 28.8 & 11.2 & + 1.2 \\
    BiGRU & $L_\textit{align}$ & 37.7 & 11.6 & + 0.0 \\
    BiGRU & $L_\textit{TS}$ ($L_\textit{CTC}$) & 25.3 & 11.2 & + 6.2 \\
    BiGRU & $L_\textit{TS+align}$ ($L_\textit{CTC}$) & 36.4 & 11.4 & + 5.1 \\
    \midrule
    UniGRU & $L_\textit{CTC}$ & 24.1 & 11.5 & + 5.3 \\
    UniGRU & $L_\textit{align}$ & 31.0 & 12.5 & + 1.8 \\
    UniGRU & $L_\textit{TS}$ ($L_\textit{CTC}$) & 22.6 & 11.9 & + 1.2 \\
    UniGRU & $L_\textit{TS}$ ($L_\textit{TS+align}$) & 32.1 & 11.5 & + 4.5 \\
    \bottomrule
\end{tabular}
\caption{Descriptive statistics about proposed models. The loss used to train the teacher model is listed in parentheses. ``\#~frames'' refers to the average number of frames with posterior greater than 0.1, ``\#~peaks'' refers to the average number of peaks in an utterance, and ``delay'' refers to the average onset of peaks in an utterance, reported relative to the average onset of peaks in the BiGRU trained with $L_\textit{align}$.}
\label{tab:stats}
\end{table}

\subsection{Teacher-Student Learning}

Another tactic for creating more useful alignments in the BiGRU is to reverse the roles of student and teacher. This strategy is based on the theory that the BiGRU will learn to produce outputs a few frames after the evidence, where a UniGRU can most confidently make a prediction. This is not likely to improve the latency, but has the potential to improve the accuracy of the UniGRU.

Teacher-student loss can be defined using $\hat{y}_{t,i}$ as the outputs of the teacher model (before softmax) at frame $i$, and $\hat{y}_{s,i}$ as the outputs of the student model at frame $i$. This loss is simply the mean-square-error between the two, with CTC loss added:

\begin{equation}
    L_\textit{TS} = L_\textit{CTC} + \frac{1}{T} \sum_{i=0}^T(\hat{y}_{t,i} - \hat{y}_{s,i})^2
\end{equation}

A comparison of the outputs of BiGRUs after training with teacher-student loss can be found in Figure \ref{fig:teacher_align}. We can see from the figure that when using the UniGRU as the teacher model, the BiGRU outputs are well-aligned with the student model. This is a stark contrast to outputs of the BiGRU trained with alignment loss. These outputs, while well-aligned with the evidence, can occur quite far before the outputs in the UniGRU, which can make it hard to learn from.

As a compromise, we train a BiGRU using both the alignment loss and the teacher-student loss, which allows the model to produce outputs useful for teacher-student learning. We can then use this better-aligned BiGRU as a teacher model for a new UniGRU student model. The full loss used to train the teacher is the sum of three losses:

\begin{equation}
    L_\textit{TS+align} = L_\textit{TS} + L_\textit{align} - L_\textit{CTC}
\end{equation}

\begin{figure}[t]
    \centering
    \includegraphics[width=\linewidth]{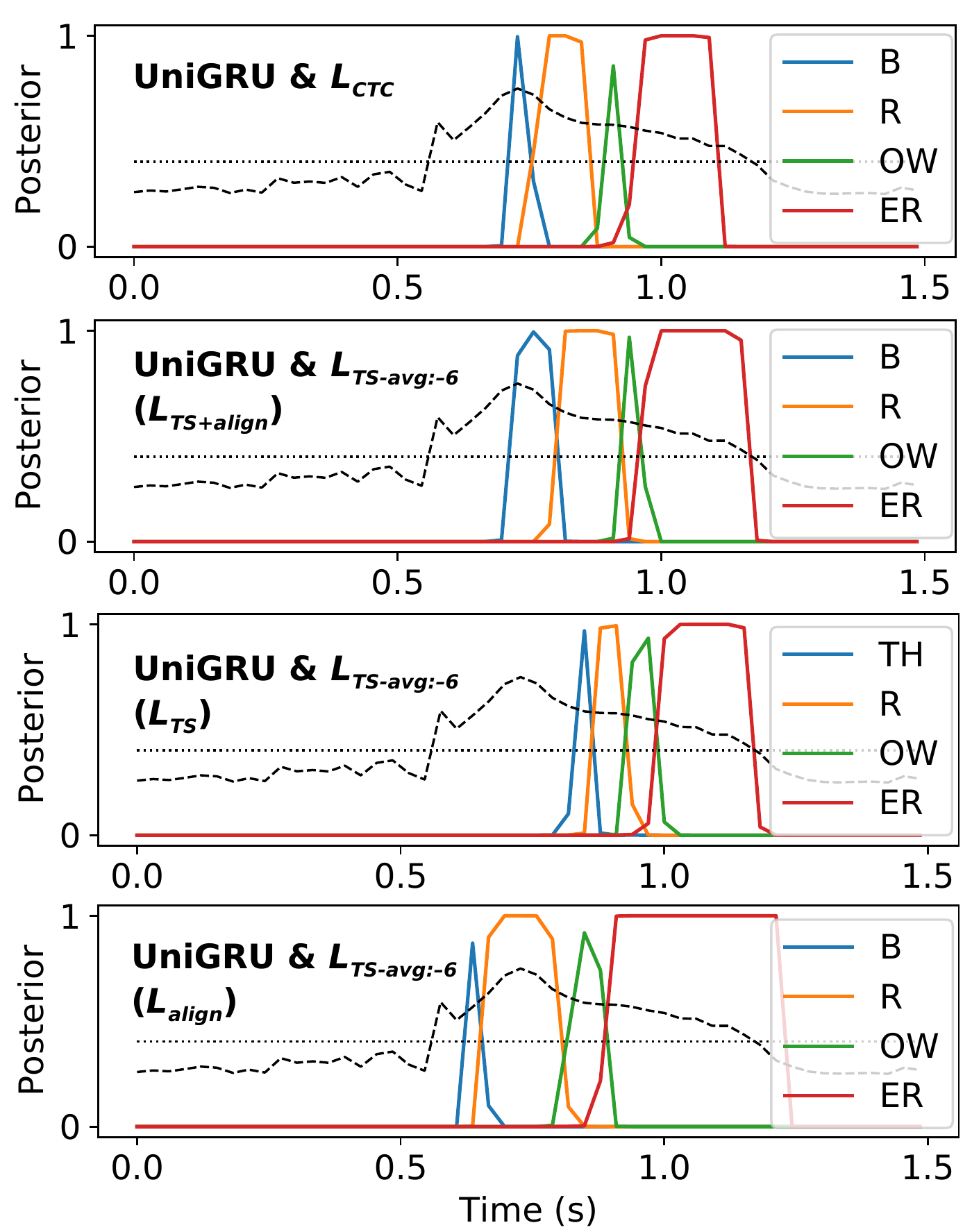}
    \caption{Comparison of alignment in UniGRUs, after training with teacher-student loss. The BiGRU is used as the teacher model, and the loss used to train it is in parentheses.}
    \label{fig:gru}
\end{figure}

We also show that we can combine our approach with the approach given by \cite{kurata2018improved} for gains in performance. This approach compares our student's test frame against not only the corresponding frame in the teacher's posterior, but also against a window of a few frames prior. There are two approaches discussed in \cite{kurata2018improved}. In one, they use only the most similar frame in the window as the target, and in the other, they use the average of frames in the window as the target. These are defined quite simply as:

\begin{equation}
    L_\textit{TS-best:N} = L_\textit{CTC} + \frac{1}{T} \sum_{i=0}^T \min_{j \in [0, N]}(\hat{y}_{t,i+j} - \hat{y}_{t,i})^2
\end{equation}

\begin{equation}
    L_\textit{TS-avg:N} = L_\textit{CTC} + \frac{1}{TN} \sum_{i=0}^T \sum_{j=0}^N(\hat{y}_{t,i+j} - \hat{y}_{t,i})^2
\end{equation}

The authors of \cite{kurata2018improved} had the best results by using the most similar frame as target, using 6 past frames as context: $L_\textit{TS-best:--6}$. However our experiments showed the average of the frames in the window provides the best target, with a trade-off between accuracy and latency depending on the width (and direction) of the window. We show the output of our systems trained using this approach in Figure \ref{fig:gru}.

\section{Experiments}

Our research is conducted with a branch of the Eesen toolkit\footnote{\url{https://github.com/srvk/eesen}} that is designed to use TensorFlow \cite{miao2015eesen}. This software has tools for extracting features and training via highly-optimized CuDNN implementation.

\subsection{Dataset}

We perform experiments using the CSLU kids' speech corpus, which is a collection of both scripted and spontaneous speech from roughly 1000 children between kindergarten and 10th grade. We focus on the scripted speech, which contains prompts of words and short phrases intended to cover a range of phonemes.

This data was not explicitly labeled with correctness of pronunciation, but as a proxy we use a label for the quality of the recording. This label can take four values:

\begin{enumerate}[noitemsep]
    \item only the target word is said
    \item target word may be present, but there's additional noise
    \item target word is not present
    \item target word is present, but there's an air puff
\end{enumerate}

We mark only the first category as "correct" and the others as "incorrect" pronunciation. It is worth noting that the data is not balanced, about 80\% of the data is marked as a 1, where the other classes each make up 5-10\% of the data.

We use 1st grade data as a test set, and 2nd grade data as a dev set. Kindergarten and 3rd through 10th grades are used as training data. To train the ASR acoustic models, we only use correctly pronounced (label 1) training samples, so that the ASR is less likely to learn to predict phones correctly when they are mispronounced. This improves model performance for the mispronunciation detection task (Table \ref{tab:md}), but increases phone error rate (Table \ref{tab:per}).

All of our experiments are done with phonemes, rather than characters, since this will allow the model to more accurately diagnose pronunciation errors. We convert each word to phonemes using the first pronunciation listed in CMUdict. For OOV words, we use CMUsphinx g2p-seq2seq\footnote{\url{https://github.com/cmusphinx/g2p-seq2seq}}. The size of the phoneme set is 39 phonemes.

\begin{table}
\centering
\begin{tabular}{llccc}
    \toprule
    Model & Loss & cPER & iPER & PER \\
    \midrule
    BiGRU & $L_\textit{CTC}$ & 5.9 & 32.6 & 12.6 \\
    BiGRU & $L_\textit{align}$ & 6.3 & 33.9 & 13.3 \\
    BiGRU & $L_\textit{TS}$ ($L_\textit{CTC}$) & 5.7 & 32.8 & 12.5 \\
    BiGRU & $L_\textit{TS+align}$ ($L_\textit{CTC}$) & 6.2 & 33.4 & 13.1 \\
    \midrule
    UniGRU & $L_\textit{CTC}$ & 12.5 & 40.2 & 19.5 \\
    UniGRU & $L_\textit{align}$ & 17.1 & 44.8 & 24.1 \\
    %GRU & $L_\textit{TS}$ ($L_\textit{TS}$) & 14.3 & 19.4 \\
    %GRU & $L_\textit{TS-best:--6}$ ($L_\textit{TS}$) & 14.1 & 19.3 \\
    UniGRU & $L_\textit{TS-avg:--6}$ ($L_\textit{CTC}$) \cite{kurata2018improved} & 14.5 & 41.5 & 21.3 \\
    UniGRU & $L_\textit{TS-avg:--6}$ ($L_\textit{TS}$) & 11.3 & 39.7 & 18.4 \\
    UniGRU & $L_\textit{TS-avg:--3}$ ($L_\textit{TS+align}$) & 11.8 & 40.3 & 19.0 \\
    \bottomrule
\end{tabular}
\caption{Phone error rate (PER) results for the CSLU kids' speech dataset. ``cPER'' stands for the error rate of correctly-pronounced words and ``iPER'' stands for the error rate of incorrectly pronounced words. Loss used to train teacher model in parentheses. Model equivalent to \cite{kurata2018improved} is marked.}
\label{tab:per}
\end{table}

\subsection{Training details}

We follow the Eesen pipeline for feature generation, which uses Kaldi software to generate standard 40-dimensional log Mel-filterbank features. We also follow the pipeline by concatenating three frames at a time into stacked frames. In the rest of this paper, a "frame" refers to one of these stacked frames.

This approach serves as a data augmentation method, since different orderings of the stack can be considered, for a three-fold increase in the size of the dataset. Stacking frames has the added benefit of reducing the memory load on the recurrent model, since utterances are one-third of their original lengths.% One last effect is that since the original frames are spaced by 10ms, the stacked frames are spaced by 30ms, slightly increasing the latency. However, most applications will find the use of stacked frames to be justified, given the significantly improved performance of the model.

For our teacher model, we use a bi-directional GRU with 4 layers of 1024 neurons (512 in each direction). In between each layer, there is an intermediate projection layer of 100 neurons, for the purpose of improving the efficiency of the model. Our student model uses 512 GRU neurons, half of the size of the teacher model, but the student model is otherwise identical. We use a dropout rate of 0.2 before and after each projection layer within the recurrent model.

We use the Adam optimizer with an initial learning rate of 0.0005, and following each epoch (starting with the 8th) halving the learning rate if the phone error rate increases. Each experiment is run for 25 epochs.

\begin{table}
\centering
\begin{tabular}{llccc}
    \toprule
    Model & Loss & P & R & F1 \\
    \midrule
    BiGRU & $L_\textit{CTC}$ & 56.0 & 64.0 & 59.7 \\
    BiGRU & $L_\textit{align}$ & 56.4 & 64.9 & 60.4 \\
    BiGRU & $L_\textit{TS}$ ($L_\textit{CTC}$) & 57.9 & 64.9 & 61.2 \\
    BiGRU & $L_\textit{TS+align}$ ($L_\textit{CTC}$) & 57.0 & 64.7 & 60.1 \\
    \midrule
    UniGRU & $L_\textit{CTC}$ & 42.5 & 76.8 & 54.7 \\
    UniGRU & $L_\textit{align}$ & 37.6 & 82.5 & 51.6 \\
    %GRU & $L_\textit{TS}$ ($L_\textit{TS}$) & 42.7 & 75.5 & 54.5 \\
    %GRU & $L_\textit{TS-best:--6}$ ($L_\textit{TS}$) & 43.2 & 74.9 & 55.1 \\
    UniGRU & $L_\textit{TS-avg:--6}$ ($L_\textit{CTC}$) \cite{kurata2018improved} & 40.1 & 78.7 & 53.1 \\
    UniGRU & $L_\textit{TS-avg:--6}$ ($L_\textit{TS}$) & 45.6 & 75.2 & 56.7 \\
    UniGRU & $L_\textit{TS-avg:--3}$ ($L_\textit{TS+align}$) & 44.9 & 75.5 & 56.2 \\
    \bottomrule
\end{tabular}
\caption{Mispronunciation detection scores for CSLU kids' speech dataset. Loss used to train the teacher model in parentheses. Model equivalent to \cite{kurata2018improved} is marked.}
\label{tab:md}
\end{table}

\section{Results and Discussion}

We first report the phone error rate (PER) in Table \ref{tab:per}. This table shows that the alignment loss tends to decrease performance for both the BiGRU and the UniGRU. The reduced performance of the BiGRU is not significant, and is probably explained by the small increase in the average number of posterior peaks (11.6 from 11.2) as seen in Table \ref{tab:stats}.

Like the BiGRU, the UniGRU also produces more peaks per utterance on average, which is likely to correspond to a small error rate increase. However, we see a larger error rate increase, which we postulate is due to the shorter delay between evidence and output. This trade-off between error rate and latency seems to be a fundamental aspect of the real-time ASR problem, and is explored more in Table \ref{tab:lag}.

For the teacher-student models, our experiments show an improvement in error rate when using the technique of \cite{kurata2018improved}. However, in contrast to their results, we find the best result from taking the average of the window of frames, rather than the best match. This may be due to the fact that the average is a "softer" target, with more information about the relative likelihoods. We find that the BiGRU trained with $L_\textit{TS}$ proves a better teacher than that used in \cite{kurata2018improved}, demonstrated by lower error rates.

In addition to the ASR evaluation, we also evaluate the ability of our model to detect mispronunciations. Since we only have utterance-level labels, we are unable to do phone-level or, in the case of phrases, word-level evaluation. Instead, we generate phone sequences with all models, and use the edit distance against the transcript (after converting to phones) to make a prediction about pronunciation correctness. We define a simple threshold PER to determine whether the pronunciation was correct. Any edit distance that is greater than 1 is marked as an incorrect pronunciation. Since the data in CSLU is not balanced, we measure performance by recording the F1 score, with mispronunciation defined as the "positive" class.

In Table \ref{tab:md}, the BiGRUs show improvement in F1 score from both the alignment loss and the teacher-student loss. For the UniGRUs, however, the alignment loss seems to hinder the performance. This is partly an artifact of the worse PER, since this will mark more pronunciations as incorrect, even though the model is already marking too many pronunciations as incorrect. This is also partly due to the earlier predictions of the UniGRU, which reduces the evidence considered by the model before making a prediction. Again, the BiGRU trained with teacher-student loss proves the best teacher.

\begin{table}
\centering
\begin{tabular}{llcc}
    \toprule
    Model & Loss & F1 & Delay \\
    \midrule
    BiGRU & $L_\textit{align}$ & 60.4 & + 0.0 \\
    BiGRU & $L_\textit{align}+L_\textit{TS}$ ($L_\textit{CTC}$) & 61.2 & + 5.1 \\
    \midrule
    UniGRU & $L_\textit{TS}$ ($L_\textit{align}$) & 47.1 & -- 0.3 \\
    UniGRU & $L_\textit{TS-avg:--6}$ ($L_\textit{align}$) & 49.1 & + 0.8 \\
    UniGRU & $L_\textit{TS-best:--6}$ ($L_\textit{align}$) & 49.5 & + 0.9 \\
    UniGRU & $L_\textit{TS}$ ($L_{CTC}$) & 51.5 & + 1.2 \\
    UniGRU & $L_\textit{TS-best:--6}$ ($L_\textit{CTC}$) \cite{kurata2018improved} & 52.1 & + 2.1 \\
    UniGRU & $L_\textit{TS-avg:--6}$ ($L_\textit{CTC}$) \cite{kurata2018improved} & 53.1 & + 3.1 \\
    UniGRU & $L_\textit{CTC}$ & 54.7 & + 5.3 \\
    UniGRU & $L_\textit{TS}$ ($L_\textit{TS}$) & 54.5 & + 5.3 \\
    UniGRU & $L_\textit{TS-best:--6}$ ($L_\textit{TS}$) & 55.1 & + 5.4 \\
    UniGRU & $L_\textit{TS-avg:--6}$ ($L_\textit{TS}$) & 56.7 & + 7.9 \\
    \midrule
    UniGRU & $L_\textit{TS-avg:+6}$ ($L_\textit{TS+align}$) & 53.2 & + 2.8 \\
    UniGRU & $L_\textit{TS-avg:+3}$ ($L_\textit{TS+align}$) & 54.5 & + 3.8 \\
    UniGRU & $L_\textit{TS}$ ($L_\textit{TS+align}$) & 55.3 & + 4.5 \\
    %GRU & $L_{TS:avg2}$ ($L_{TS+align}$) & 55.8 & + 3.9 \\
    UniGRU & $L_\textit{TS-avg:--3}$ ($L_\textit{TS+align}$) & 56.2 & + 5.4 \\
    %GRU & $L_{TS:avg4}$ ($L_{TS+align}$) & 55.5 & + 4.5 \\
    UniGRU & $L_\textit{TS-avg:--6}$ ($L_\textit{TS+align}$) & 55.5 & + 5.9 \\
    \bottomrule
\end{tabular}
\caption{Results demonstrating the trade-off between accuracy and latency. Loss used to train teacher model is in parentheses. Models equivalent to  \cite{kurata2018improved} are marked.}
\label{tab:lag}
\end{table}

Finally, we show in Table \ref{tab:lag} that there is a trade-off to be made between accuracy and latency. The BiGRU trained with $L_\textit{align}$ provides our best estimate of the true onsets, so we list our other results as a comparison to those results. Since each frame lasts 30 ms, our results range in latency from -10 ms to 240 ms.

The table includes a list of models trained with teacher-student training in order of increasing latency. The F1 score closely follows the trend of the latency, increasing as the latency increases. This clearly demonstrates the trade-off between latency and performance.

In the last few rows of the table, we show results from our proposed teacher model, the BiGRU trained with $L_\textit{TS+align}$. We experiment with both past and future windows of frames from the teacher model, and find a similar trade-off between accuracy and latency. Compared to the other models listed, this model has better latency than models with comparable accuracy, and better accuracy than models with comparable latency.

\section{Conclusion}

We have taken a well-performing BiGRU, and gleaned important cues for use by a smaller and real-time appropriate UniGRU. In order to accomplish this, we developed an "alignment loss" that produces well-aligned outputs in the BiGRU and combine this with teacher-student training with the UniGRU as the teacher. The well-aligned BiGRU allows us to effectively train a UniGRU. We can tweak the size and direction of a window of frames to trade off between mispronunciation detection accuracy and latency.

We use a fairly simple model for mispronunciation in this work, that does not distinguish between errors made by the model and errors made in pronunciation. We hope to address this in the future by using error modeling, distinguishing ASR errors from pronunciation errors.

Another direction for future work is to develop strategies for continuous speech, including finding word boundaries, ignoring pauses, and locating the position of the utterance in the text if the reader skips words.

\bibliographystyle{IEEEbib}

\bibliography{mybib}

\end{document}